\documentclass[pdflatex,sn-mathphys-num]{sn-jnl}

\usepackage{graphicx}%
\usepackage{multirow}%
\usepackage{amsmath,amssymb,amsfonts}%
\usepackage{amsthm}%
\usepackage{mathrsfs}%
\usepackage[title]{appendix}%
\usepackage{xcolor}%
\usepackage{textcomp}%
\usepackage{manyfoot}%
\usepackage{booktabs}%
\usepackage{algorithm}%
\usepackage{algorithmicx}%
\usepackage{algpseudocode}%
\usepackage{listings}%

\graphicspath{{./}{figures/}}

\newcommand{\be}{\begin{equation}}
\newcommand{\ee}{\end{equation}}
\newcommand{\bea}{\begin{eqnarray}}
\newcommand{\eea}{\end{eqnarray}}

\newcommand{\lapprox} {\, \lower3pt\hbox{$\sim$}\llap{\raise2pt\hbox{$<$}}\,}
\newcommand{\gapprox} {\, \lower3pt\hbox{$\sim$}\llap{\raise2pt\hbox{$>$}}\,}

\begin{document}

\title [Expanding Cosmological Singular Horizon]{An  Event Horizon `Firewall' Undergoing Cosmological Expansion}

\author[1]{\fnm{R. N.} \sur{Henriksen}}\email{henrikr@queensu.ca}

\author[2]{\fnm{A. G.} \sur{Emslie}}\email{gordon.emslie@wku.edu}

\affil[1]{\orgname{Queen's University}, \orgaddress{\street{Stirling Hall, Bader Crescent}, \city{Kingston}, \postcode{ON K7L 3N6}, \country{Canada}}}

\affil[2]{\orgdiv{Department of Physics \& Astronomy}, \orgname{Western Kentucky University}, \city{Bowling Green}, \postcode{KY 42101}, \country{USA}}


\abstract{We embed an object with a singular horizon structure, reminiscent of (but fundamentally different from, except in a limiting case) a black-hole event horizon, in an expanding, spherically symmetric, homogeneous, Universe that has a positive cosmological constant. Conformal representation is discussed. There is a temperature/pressure singularity and a corresponding scalar curvature singularity at the horizon. The expanding singular horizon ultimately bounds the entire space-time manifold. It is is preceded by an expanding light front, which separates the space-time affected by the singularity from that which is not yet affected. An appropriately located observer in front of the light front can have a Hubble-Lema\^itre constant that is consistent with that currently observed.}

\keywords{cosmology, self-similarity}



\maketitle


\section{Introduction}

Recently, in \cite{FCZT2023,CFPT2023}, the importance of embedding a black hole type singularity in an expanding cosmology has been emphasized; the resulting black hole would grow with the Universe, rather than remain stationary in asymptotic Minkowski space.  Ideally, one would like a locally rotating, perhaps non-stationary, Kerr-type solution that asymptotically passes smoothly into a part of, or even all of, an expanding space-time. Such a solution is not known, and a successful one might have to be embedded in a universe that is not only expanding, but is also itself rotating. Even in spherical symmetry, the best attempt at coupling a local black hole smoothly to a universe that expands to infinity, follows from \cite{McV1933} and its conformal descriptions \citep{LA2011}.

We are unable to deliver this most desirable embedded solution. The more modest purpose of this work is to recall an earlier cosmological solution \citep{HEW1983,CH1991,H2015}, which, for a different parameter range than that originally considered, also allows an interpretation in terms of an object with a singular horizon that is embedded in an expanding cosmology. 

It is a solution to the Einstein field equations that possesses non-stationary spherical symmetry (with zero shear and isotropic expansion) and an ideal fluid matter description \citep{M1969}. It contains the cosmological constant as the simplest model of vacuum energy. 

Despite the existence of the cosmological constant, the field equations may nevertheless be solved explicitly as a self-similar (i.e., kinematic self-similarity, \citep{CH1991}) form. Such a self-similar solution is the smoothest way of coupling a solution from small to large scales. The solution fits into the ansatz prescribed in \cite{McV1967} for finding homogeneous solutions; however, this particular example was found independently.

By requiring this self-similar symmetry, a particular equation of state is imposed on the cosmology. The resulting form of the equation of state is not unreasonable in the hot gas phase of the cosmology and it may pass smoothly to the gravitational collapse phase.

In contrast to the cases explored in \cite{HEW1983}, the solutions discussed here have an a horizon that is a null surface (zero light speed) and defines the boundary of an embedded object. On a (necessarily) cosmic time scale, the radius coordinate of this surface expands exponentially as a function of coordinate time. 

We may somewhat loosely refer to the central object within the  horizon as a ``black hole'' in the general case, because light cannot traverse the horizon. However the structure formally differs from a Schwarzchild or Kerr black hole, because the pressure, and consequently the scalar curvature, are infinite at the horizon. It is as though the central singularity has been ``pasted'' onto the horizon surface, rather than being located at the center of the embedded object.

 For understanding any physical application of the solution it is necessary to appreciate the very large scales associated with the time and space variables. These scales (see Equations~(\ref{fundamental-scale-length}) and~(\ref{fundamental-scale-time}) below) are set by the fundamental constants $c$ (speed of light) and $\Lambda$ (cosmological constant), and they correspond to spans of order the life time and size of the Universe, respectively. The Hubble-Lema\^itre expansion rate varies from a minimum value, given in terms of fundamental constants, to values that are familiar to observers at the current epoch, allowing us to straightforwardly locate an observer within the space-time solution.

We do not present our solution as an wholly explored cosmological model, although far from the singularity it is rather close to the $\Lambda-{\rm CDM}$ model. It is moreover not the theoretical solution to embedding a rotating black hole in an expanding cosmology. Our solution does, however, add to the list of previous works aimed at embedding an object shielded by a horizon within an expanding universe, in what we believe is an original way \citep[cf.][]{CH1991,H2015}. We elaborate at some length on the local physical behavior of the cosmology, because this solution may indeed have some relevance to recent cosmological observations \citep{FCZT2023}.

In the next section we review the solution of \cite{HEW1983} and we derive the correspondence between physical quantities \citep{M1969} and the quantities that we use in describing the solution, in particular the pivotal parameter $C$. Subsequent sections present details of the solution and discuss its interpretation.

\section{Physical Quantities}\label{sect:definitions}

The solution to the general relativity field equations in \cite{HEW1983} is given in terms of a combined space-time variable
\be\label{eq:xi}
\xi = \frac{\sqrt{3/\Lambda}}{r} \, \exp \left ( \sqrt{\frac{\Lambda c^2}{3}}~t \right ) \,\,\, ,
\ee
where $\Lambda$ is the cosmological constant, $c$ is the speed of light in vacuo, and $t$ and $r$ are convenient time and space coordinates. The obvious scale for any space measure is
\be\label{fundamental-scale-length}
\sqrt{3/\Lambda} \approx 1.65 \times 10^{28} \, {\rm cm} 
\ee
(where we have adopted the Planck measurement $\Lambda \approx 1.11 \times 10^{-56} \, {\rm cm}^{-2}$), while the corresponding time scale is
\be\label{fundamental-scale-time}
\frac{\sqrt{3/\Lambda}}{c} \approx 5.5 \times 10^{17} \, {\rm s} \approx 1.75 \times 10^{10} \, {\rm years} \,\,\, .
\ee
With these units for $r$ and $t$, the self-similar space-time variable becomes simply
\be\label{eq:xiscaled}
\xi=\frac{e^t}{r} \,\,\, ,
\ee
where it is recognized that changes of order unity in $t$ or $r$ span the respective scales of the known Universe.

In this example of kinematic self-similarity \citep{CH1991} \cite{H2015}, the metric of spherically symmetric space time takes the form
\be\label{eq:metric}
ds^2 = e^{\sigma(\xi)} \, dt^2 - e^{\omega(\xi)} \, dr^2 - R^2(\xi) \, (d\theta^2 + \sin^2\theta \, d\phi^2) \,\,\, , 
\ee
where the circumferential radius
\be\label{eq:radiius}
R(r,\xi) = r \, S(\xi) \,\,\, ,
\ee
so that, at a given spatial coordinate $r$, the function $S(\xi)$ plays the role of the scale factor $a(t)$ in the standard FLRW cosmology.

The matter mass $m_m$ is the total mass $m$ corrected by subtracting the mass of the vacuum:
\be\label{eq:mass}
m_m= m - \frac{4\pi}{3} \, R^3 \, \rho_v = \frac{r}{2} \, M(\xi) \,\,\, ,
\ee
where
\be\label{eq:rho-vacuum-physical-value}
\rho_v = \frac{\Lambda \, c^2}{8\pi G}\approx 5.89 \times 10^{-30} \, {\rm g}~{\rm cm}^{-3}
\ee
and $G$ is the Newtonian gravitational constant. Each mass term in Equation~(\ref{eq:mass}) is dimensionless, related to the physical mass $m_m$ (or to the vacuum mass $m_v = 4\pi R^3 \, \rho_v/3$; each in grams) by the factor
\be\label{eq:value_of_mu}
\mu = \sqrt{\frac{3}{\Lambda}} \, \frac{c^2}{G} ~\approx 2.26 \times 10^{56} \, {\rm g} \,\,\, ;
\ee
the dimensional masses are thus of order of the mass of the observable Universe.

The matter density $\rho_m$ and the matter pressure $p_m$ are respectively given by
\be\label{eq:mdensmpress}
8 \pi \rho_m = \frac{\eta(\xi)}{r^2} \, ; \qquad 8 \pi p_m = \frac{P(\xi)}{r^2} \,\,\, .
\ee
The units of $\eta(\xi)$ may be taken to be in terns of $\rho_v$, while the units of $P(\xi)$ are $\rho_v \, c^2 \approx 5.27 \times 10^{-9}$~erg~cm$^{-3}$. In each case $r$ is the dimensionless radius.

In \cite{HEW1983}, the Einstein field equations for a spherically symmetric, non-stationary, space-time manifold were used, together with the Bianchi identities \citep{M1969}, to write a complete set of ordinary differential equations for $S(\xi)$, $M(\xi)$, $P(\xi)$, $\eta(\xi)$, $\sigma(\xi)$ and $\omega(\xi)$. For this the authors had to show that self-similarity, in terms of $\xi$ as the invariant, existed in the equations. Without having to assign an equation of state \emph{a priori}, this completeness comes from the splitting of the energy equation in a manner that achieves the desired self-similarity.

It is therefore useful to recall the energy equation in the form
\be\label{eq:energy}
\Gamma^2 - \frac{U^2}{c^2} = 1-\frac{2 \, G \, m}{R \, c^2} \,\,\, ,
\ee
where $m$ is the total mass, the co-moving Lorentz factor
\be\label{eq:Gamma}
\Gamma = e^{-\omega/2} \, \partial_r R \,\,\, ,
\ee
and the radial four velocity of a `comoving observer' (i.e., one at fixed $r$) is (in units of $c$)
\be\label{eq:fourvel}
U = e^{-\sigma/2} \, \partial_t R \, \equiv r \times e^{-\sigma/2} \, \xi S'(\xi) \,\,\, .
\ee

The energy equation~(\ref{eq:energy}) is the equation that must be split in two, as in the original paper \cite{HEW1983}.  The separation is a simplifying assumption, different from those normally used (e.g., \cite{KSMH1980}, page~167). This procedure splits the matter and vacuum mass terms according to
\bea
\Gamma^2(\xi) & \equiv & e^{-\omega(\xi)} \, (S(\xi) - \xi S'(\xi))^2 \cr
& = & 1-\frac{2 \, G \, m_m}{R \, c^2} \equiv 1-\frac{M(\xi)}{S(\xi)} \,\,\, ,
\eea
and
\be\label{eq:sigma1}
U^2 \equiv r^2 e^{-\sigma} \, (\xi S'(\xi))^2 = \frac{2 \, G \, m_v}{R \, c^2} \equiv r^2 S^2(\xi)=R^2\,\,\, ,
\ee
where the prime indicates differentiation with respect to $\xi$, and the expressions are dimensionless. Given the sign structure in the energy equation, this separation is unique for a positive cosmological constant.

\section{Solution} \label{sec:solution}

As shown in \cite{HEW1983}, there is a general integral of the field equations that takes the form
\be\label{eq:integral}
M(\xi) = \eta(\xi) \, \frac{S^3(\xi)}{3} + \Delta \, ,
\ee
where $\Delta$ is a constant. When $\Delta \ne 0$, a numerical solution is possible, but we focus here on the analytic solution that results when $\Delta=0$. In terms of the quantities used in the previous section, the solution takes the form \citep{HEW1983}
\bea\label{eq:cosmosol}
 e^{\sigma} \!\! = & (C \tanh(C \, \ln \xi) - 1)^2,  \,\, e^{\omega} = & C^2 \, K^2 \, \xi^2 \, {\rm sech}^2 (C \, \ln \xi)  ,\cr \cr
S  = & \,\, K \, \xi \, {\rm sech} (C\ln \xi) \,\,\,\,\,\,\,\,\,\,\,\,\, , \,\, M = & K \, \xi \, {\rm sech}^3 (C \, \ln \xi) \, , \cr \cr
P = & \frac{1 - 3 \, C \tanh(C \, \ln \xi) }{K^2 \xi^2 (C \tanh(C \, \ln \xi) - 1)} \,\,\,\, , \,\,\,\,\, \eta \,\, = & \frac{3}{K^2 \, \xi^2} \,\,\, .
\eea

Noting that $CS = e^{\omega/2}$ and hence that $R = r\,S=r\,e^{\omega/2}/C$, it follows 
(\cite{KSMH1980}, page~166) that the shear is zero.
The expansion is finite:
\be
\Theta = 3 \, e^{-\sigma/2} (1 - C \,\tanh(C \ln{\xi})) = 3 
\ee
and is isotropic. 

The solution fits into the McVittie \cite{McV1967} `ansatz' as presented in section 14.2.3 in \cite{KSMH1980}, although as far as we know this particular possibility has not been explored elsewhere. Therefore, for classification purposes we next demonstrate this briefly. 

Referring to the notation in (\cite{KSMH1980}) we have $\lambda=\omega/2$, $\nu=\sigma/2$ in our terminology. Then, as quoted in \cite{KSMH1980}, the ansatz requires 

\be
e^{\omega/2} = P(r) \, S(t) \,e^{\eta(z)},
\ee
where $z=\ln{(Q(r)/S(t))}$, and $S(t)$ is not the same as our $S(\xi)$. We can choose $S=e^t$, $P(r)=CK/r$ and $Q(r)=r$. This gives $z=-\ln{\xi}$, so that taking 

\be
\eta (-\ln{\xi})=\ln{( {\rm sech}(C \ln{\xi}))},
\ee
we have 

\be
e^{\omega/2} = C \, K \, \xi ({\rm sech}(C\ln \xi))
\ee
as given in equation~(\ref{eq:cosmosol}).

With $\dot S/S=1$, the ansatz requires the quantity $f(t)=0$ and so (equation 14.25 in \cite{KSMH1980})  
\be
\frac{\sigma}{2}= \ln \left ( \frac{\dot\omega}{2} \right ) \,\,\, .\label{eq:sigmaomega}
\ee
Using the expression for $e^{\omega/2}$ in equation (\ref{eq:cosmosol}) gives
\be
\frac{\dot{\omega}}{2} = \frac{d \ln [ \xi \, {\rm sech} (C \, \ln \xi) ]}{d \ln \xi}  = 1 - C \, {\rm tanh} (C \, \ln \xi) \,\,\, ,
\ee
so that 
\be
\frac{\sigma}{2}=\ln{(1-C\tanh{(C\ln{\xi}}))} \,\,\, ,
\ee
as in equation~(\ref{eq:cosmosol}).

Our metric also fits the general form for homogeneous solutions (equation 14.44 in \cite{KSMH1980}) because $f(t)=0$ and equation (\ref{eq:sigmaomega}) holds. Minor re-scalings and re-definitions are required.

Using the remaining field equations, expressions for the pressure and density can be found in terms of this ansatz. However, having split the energy equation, we have diverged from this procedure and are instead led to the analytic solution~(\ref{eq:cosmosol}).

The constant $K$ is not without physical significance because it does not appear together with $\xi$ in the arguments of the hyperbolic functions (otherwise it could be absorbed into $\xi$). It is convenient to keep this freedom explicit in the formal expressions, but we will ignore it in our numerical estimates. (It will be shown below that $K = 1/C \simeq 1$ in the most likely  case.) The constant $C$ is the principal physically significant constant, and because ${\rm sech} \, x$ and $x \tanh x$ are both even functions, it may be taken to be positive without loss of generality.

Given the form of $g_{00} = e^{\sigma}$, there is no horizon if $C<1$ (these cases formed the basis for the work in \cite{HEW1983}). However, when $C>1$, $e^\sigma \rightarrow 0$ at $\xi_h$, where $\tanh(C\ln\xi_h) = 1/C$, or
\be\label{eq:xih}
\xi_h = \left( \frac{C+1} {C-1} \right)^{1/2C} \,\,\, .
\ee

We note that $\xi_h$ is called here a `horizon' because the radial null velocity
\be\label{eq:nullsurface}
\frac{dr}{dt} = e^{\sigma/2} \, e^{-\omega/2}
\ee
equals zero at $\xi=\xi_h$. However at the horizon, $C \, \tanh(C \, \ln \xi) - 1 = 0$ and so, from Equations~(\ref{eq:cosmosol}), $P \rightarrow -\infty$. Hence, via the contracted field equations, the Ricci curvature invariant $\Re$ also becomes infinite at the horizon.

If the quantity $g^{\mu\nu}F_{,\mu}F_{,\nu}=e^{-\sigma} (\partial_t F)^2-e^{-\omega}(\partial_r F)^2 $, with $F=e^\sigma$, is null at $\xi = \xi_h$, then the horizon is a null surface. A straightforward calculation shows, however, that the normal to the surface is \emph{not} null at $\xi = \xi_h$, but rather is finite and time like. It does approach zero as $C\rightarrow 1$.

The question arises as to whether the singularity is `naked.' To this end we have numerically integrated the explicit equation for the null paths in $(t,r)$ coordinates, viz. (with $K=1/C$):
\bea
\frac{dr}{dt} & = & \mp \, \frac{C \, {\rm tanh} (C \ln \xi) - 1}{\xi \, {\rm sech} (C \ln \xi)} \cr
& = & \mp\limsup{} \, \frac{1}{\xi} \, [ C \sinh (C \ln \xi) - \cosh (C \ln \xi)] \cr
& = &\mp \limsup{} \, \frac{1}{2 \, \xi} \, \left [ (C - 1) \, \xi^C - (C + 1) \, \xi^{-C} \right ] \cr
& = & \mp \, \frac{r}{2 \, e^t} \, \left [  (C - 1) \, \frac{e^{Ct}}{r^C} - (C + 1) \, \frac{r^C}{e^{Ct}} \, \right ] \,\,\, ,
\label{eq:null_path_0}
\eea
for various values of $C \ge 1$. 

We started the integration from an initial time $t=\ln{(r \, \xi_h})$ for various values of $r$, thus determining $r(t)$ for light rays that end at the singularity.

In Figure~\ref{fig:conformal} (see Appendix) we sketch schematic conformal diagrams corresponding to these null ray solutions for the three cases: $C < 1$ (the original cosmology in \cite{HEW1983}), $C=1$ (the singular case described in Section~\ref{sect:LC} below), and $C > 1$. 
In section~\ref{sect:LC} we show in some detail approximations to the null geodesics that justify our sketches.  
In these schematic diagrams, the $(t,r)$ coordinates are `fluid' in the sense that light rays are not necessarily at $45^\circ$, as would be usual. \footnote{We also show a ``strict'' conformal diagram for the case $C=1$ in Figure~\ref{fig:strconform} in the Appendix. This uses the exact solution of equation (\ref{eq:null_path_0}) available when $C=1$.} 
On the other hand the shape of the singularity in these diagrams is reasonably accurate. Unlike FLRW models, the time line $r=0$ does have some physical significance, because the pressure is not homogeneous.

 At large negative $t$ it is the second term in equation (\ref{eq:null_path_0}) that dictates the behavior of the null geodesic. This gives for the outgoing rays
 \be
 r^C=\frac{2}{Ce^{-(C+1)t}+C\,v},
 \ee
 where $v$ is constant on the ray. This shows the affine parameter $t$ extending ultimately to $i^-$ as $r\rightarrow 0$, with the rays being released at various values of $v$ en route.
 At negative $t$ one can find the time for a given ray when $r=e^t/\xi_h$ using this approximation (and also at positive $t$ using the first term in equation (\ref{eq:null_path_0})), but numerically 
 we find (see discussion in section~\ref{sect:LC} that all rays emitted back to past time-like infinity $i^-$  are absorbed by the singularity before reaching spatial infinity $i_o$). 
 
 In the limiting case with $C=1$ precisely, the singularity is also not naked; in this case, in $(t,r)$ coordinates, the singularity is time-like at finite times but space-like at the end of time where the outgoing light rays arrive according to the exact expression $r=1/(v+e^{-2t})$ ($v$ constant).
 
\section{Physical Implications of the Solution}\label{sect:Cphys}

The importance of the parameter $C$ justifies finding an expression that relates it to physical quantities.

Equations~(\ref{eq:cosmosol}) \citep[see also Equations~(34), (43) and~(46) of][]{HEW1983} relate the constant $C$ to the values of the dimensionless parameters $P$, $\eta$, $M$, and $S$:
\be\label{eq:C-alpha-M-S}
P = \eta \, \frac { C \sqrt{1 - M/S} - \frac{1}{3}} {1 - C \sqrt {1 - M/S} } \,\,\, .
\ee
Solving this for $C$:
\be\label{eq:C-solution}
C = \frac{3 P + \eta}{ 3 \, ( P + \eta) }  \frac {1} {\sqrt{1 - M/S}} \,\,\, .
\ee
Generally, if we assign numerical values to $P, \eta, M$ and $S$ to obtain $C$, then the value of $\xi$ at the point in question is then fixed by any one of  Equations~(\ref{eq:cosmosol}). In terms of dimensional physical quantities, equation~(\ref{eq:C-solution}) is
\be\label{eq:physicalC}
C = \frac{3 \, p_m + \rho_m}{3 \, (p_m + \rho_m) }  \frac {1} { \sqrt{1 - \frac{2 m_m}{R} }} \equiv \frac {\beta} {\sqrt{1 - \frac{2 m_m}{R}} } \,\,\, ,
\ee
where we have defined
\be\label{eq:beta-definition}
\beta = \frac{3 \, p_m + \rho_m}{3 \, (p_m + \rho_m) } \,\,\, .
\ee
From Equations~(\ref{eq:cosmosol}), $e^{\omega/2} = C S$, so the spatial part of the metric~(\ref{eq:metric}) becomes
\bea\label{eq:metric-spatial-part}
C^2 \, S^2 dr^2 & + & r^2 \, S^2 \, d\Omega^2 = S^2 \left [ C^2 dr^2 + r^2 d\Omega^2 \right ] \cr
& = & S^2 \left [ \beta^2 \, \frac{dr^2}{1 - \frac{2 m_m}{R}} + r^2 \, d\Omega^2 \right ] \,\,\, ,
\eea
where we have used Equation~(\ref{eq:radiius}). With $\rho_m = 0$, as at infinite time, $\beta = 1$ and the spatial part of the metric becomes an ``expanding Schwarzchild'' form $S^2 \, [ dr^2/(1 - 2 m_m/R) + r^2 \, d\Omega^2 ]$. Should $\rho_m + 3 p_m = 0$ (which is $P=-\eta/3$), as when $\xi=1$ (see  below), then $\beta = 0$ and the spatial part of metric reduces to the purely circumferential part $S^2 \, r^2 \, d\Omega^2$, that is a `two sphere'.

The functions in the variable $\xi$  in equations (\ref{eq:cosmosol}) are well behaved but for two incidents: the previously-noted  singularity in  $P$ at $\xi = \xi_h$, and also singularities in both $\eta$ and $P$ where $r \rightarrow \infty$ (or $t \rightarrow -\infty$), both corresponding to $\xi \rightarrow 0$. In fact, $P$ approaches negative infinity as $\xi\rightarrow 0$ for any value of $C$.
Moreover, when $C>1$, $P$ can also become negative at large $\xi$, although the horizon can be set so that this negative value is hidden. 

We emphasize that apart from these two incidents, the expressions in Equations~(\ref{eq:cosmosol}) give the solution formally everywhere, both outside and inside the horizon. This would allow us to speculate on properties inside the apparent horizon, were it not for the pressure singularity there.  

We also note that \emph{there is no singularity at $r=0$}: as $r \rightarrow 0$, $\xi \rightarrow \infty$ (Equation~(\ref{eq:xiscaled})), so that, with $C > 1$, Equations~(\ref{eq:cosmosol}) cause $R (= rS)$, $e^\omega $, $m_m (= r M)$ to approach zero. Moreover $\rho_m = \eta/(8\pi r^2)$ and $p_m = P/(8\pi r^2)$ are finite at finite $t$ (i.e., not at $t^-$). The metric coefficient $g_{00} = e^\sigma$ approaches the value $(C-1)^2$ so that initially $ds>0$ at $i^-$. 

The singularity that appears inside the horizon in, (e.g., the Schwarzschild black-hole solution), is thus displaced from the origin to the location of the horizon itself, In this sense the solution fundamentally differs from that for an embedded black hole. It describes the future evolution of an initial `bubble' of vacuum at $i^-$, which has a thermal singularity on the boundary.

The ``wave front'' $\xi=1$ is another significant point in the solution: at this (propagating) location $r = e^t$, $M = S = 1$, $P = -1$, $\eta=3$, $e^\sigma = 1$, $e^\omega = C^2$, and $\Gamma^2 = (1/C) (S-\xi S') = 0$. Moreover $3 p_m + \rho_m \propto (3 P + \eta) = 0$ at this point, and Equation~(\ref{eq:C-solution})  shows that this is only possible when $M=S$, i.e., $2 G\, m_m/c^2 = R$. We also note that $S'(\xi)=K$ at $\xi=1$. The latter result implies that $R$ is a maximum at any finite $r$ when $t = \ln{( r)}$. 
At the horizon $\xi_h$, $S'(\xi_h) = 0$  (from Equation~(\ref{eq:cosmosol})). 

A cosmological application of the solution requires returning to ($t,r$) variables, which we will pursue briefly in the next section. However, we recall that the relation between $\eta(\xi)$ and  $\rho_m$  reminds us that the solution is homogeneous as well as spherically symmetric and non stationary.  This is true even though metric quantities, the mass, and the pressure do vary with radius at any finite time. The system center is the center of the embedded object when $C>1$.

The general behavior of physical characteristics of the solution is similar whether there is a horizon or not, except that, absent the singular horizon that occurs for $C \ge 1$, the central singularity ($\xi=0$) is not hidden.

\section{The Cosmology in Co-moving Coordinates}\label{sec:interpretation}

 The singular horizon is at a fixed value of $\xi=\xi_h$; so that at a fixed time there is a minimum accessible radius, and at a fixed $r$ there is a maximum accessible future time. This is as ``seen'' by an observer at $\xi_o$ for whom $\xi_o<\xi_h$.

A significant issue for the cosmology lies in appropriately situating an observer $\xi_o$ in the solution. A useful reference value is the value of the Hubble-Lema\^itre expansion rate. This is given in various forms as
\be\label{eq:H1}
H\equiv \frac{e^{-\sigma/2} R_t}{R} = e^{-\sigma/2} \, \frac{\xi \, S'(\xi)}{S}= \frac{U(\xi)}{R}=1 \,\,\, ,
\ee
where the last expression follows from Equation~(\ref{eq:sigma1}). The proper value is therefore constant. The unit is that of reciprocal time, and the pertinent scale is  $1/(5.51\times 10^{17})$ sec$^{-1}$, which is equal to $56$~km~s$^{-1}$~Mpc$^{-1}$. This does not exactly locate our epoch in our cosmology, but it is close.

In calculating this value of $H$ we have used proper time in the cosmological metric. In the FLRW cosmologies the proper time is the coordinate time. If we use the coordinate time in our cosmology to calculate $H(\xi)=R_t/R$, we find that
\be
 H(\xi)=\frac{\xi S'(\xi)}{S(\xi)} = 1 - C \, \tanh(C \, \ln{\xi}) \, \equiv \, e^{\sigma/2} \,\,\, ,\label{eq:Ht}
\ee
in units of $56$~km~s$^{-1}$~Mpc$^{-1}$. This value is smaller than unity at $\xi>1$ and larger than unity at $\xi<1$. It is again equal to $1$ at $\xi=1$.

At the apparent horizon $\xi_h$, the value of `coordinate' $H$ is zero. A good place to locate a terrestrial observer is therefore at $\xi_o = 0.78$, because $H(0.78) \approx 1.25$, equivalent to $70$~km~s$^{-1}$~Mpc$^{-1}$, and so in approximate agreement with the (somewhat varying) observational inferences. 

In the physical discussion of this section we will (arbitrarily) take $C$ to have a value just above the threshold value $C=1$. Specifically, we shall take $C=1.01$, which places the horizon singularity at $\xi_h = 201^{1/2.02} \simeq 13.8$ (Equation~(\ref{eq:xih})). The observer is located at $\xi_o = 0.78$ (see above), which is comfortably distant from the horizon.

For the two key values of $\xi$ (the horizon $\xi_h$ and the observer $\xi_o$), there are two corresponding ``waves'' that we can follow:
\be\label{eq:specialR}
r_h = \frac{e^t}{\xi_h} \, ; \qquad r_o = \frac{e^t}{\xi_o} \,\,\, ,
\ee
giving the coordinates $r$ at which the corresponding surfaces are found at coordinate time $t$. We will present quantitative results for two illustrative values of $t$, specifically:
\begin{itemize}
    \item $t=1$, when the horizon is at $r_h = e^1/13.8 = 0.197$ and the observer is at $r_o = e^1/0.78 = 3.48$;

    \vskip 0.2in
    
    \item $t=0.1$ (Figure~\ref{fig:HEWR}), for which $r_h = e^{0.1}/13.8 = 0.008$ and $r_o = e^{0.1}/0.78 = 1.42$.
\end{itemize}
We can also treat these ``waves'' in terms of the coordinate time at which they pass a given location $r$, viz.
\be\label{eq:specialT}
t_h = \ln(\xi_h \, r) \, ; \qquad t_o = \ln(\xi_o \, r) \,\,\, ,
\ee
and we shall present results for two illustrative values of $r$, viz.

\begin{itemize}
    \item $r=1$, where the horizon passes at time $t_h = \ln(13.8) = 2.62$ and the observer passes at $t_o = \ln (0.78) = -0.25$;

    \vskip 0.2in
    
    \item $r=2$ (Figure~\ref{fig:HEWM}), where the singularity  arrives at $t_h=\ln(27.6) = 3.32$ and the observer passes at $t_o=\ln(1.56)=0.445$. 
\end{itemize}

The resulting behaviors that we show below are qualitatively typical of those for any value $C>1$, but with quantitative changes in the time and space positions of interesting features. (A value $C=0.99$ gives a similar cosmology without an apparent horizon and so without a singular horizon.) We will set the constant $K=1$ for the illustrations; in general we expect $K = 1/C$ (see section~\ref{sec:lorentz-factor}).

\subsection{Circumferential Radius}

\begin{figure}
\includegraphics[width=0.99\linewidth]{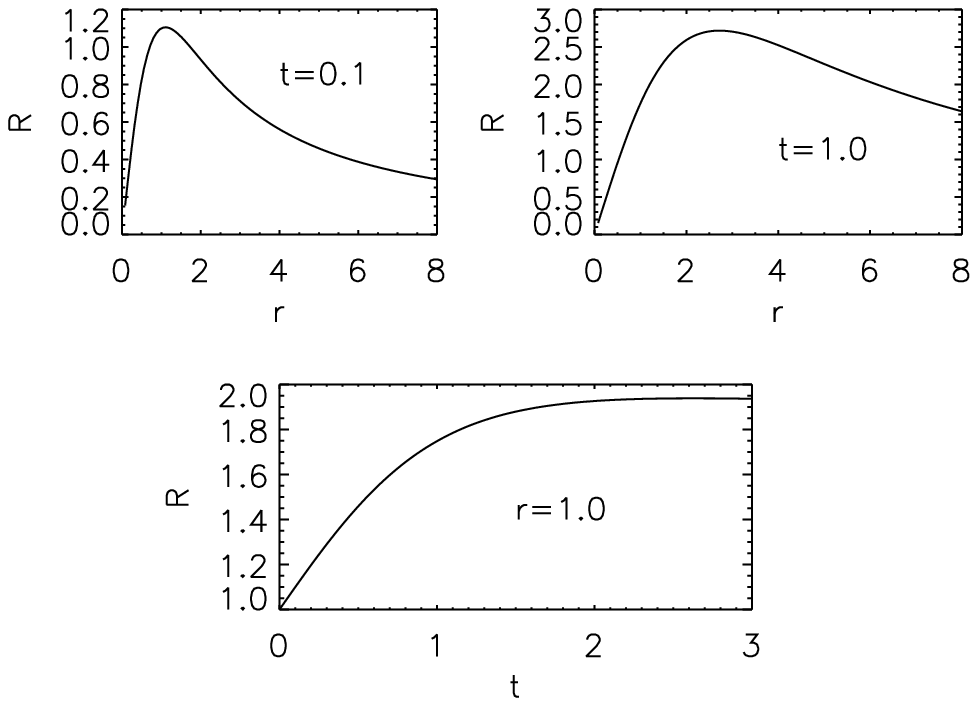}
\caption{At the upper left we show the circumferential radius as a function of $r$ at $t=0.1$. (This places the singularity radius at $r_h=0.08$). With $S_h(13.81)=1.94$, the circumferential radius is $R_h=0.155$. At upper right we plot the same quantity for a later time $t=1$. The singularity radius is now at $r_h=0.197$, which, when multiplied by $S(\xi_h)$, gives a circumferential radius $R_h = 0.382$. The maximum value of $R$ occurs at $\xi=1$ in each case, e.g., at a value of $r = e^t \simeq 1.1$ on the left; the value of the maximum also equals $e^t$. On the left the observer (who ``sees'' the Hubble-Lema\^itre quantity as $70$~km~s$^{-1}$~Mpc$^{-1}$, i.e, $H=1.25$ in the units used here) is at $\xi = 0.78$, $r=1.41$, and therefore $S(0.78) = 0.756$ and $R = 1.07$. The lower figure shows the circumferential radius as a function of $t$ at $r=1$. It has a rather broad maximum value $\simeq 2$ that occurs at a time $t \simeq \ln (\sqrt{200}) \simeq 2.65$.} 
\label{fig:HEWR}
\end{figure}

The top panels of Figure~\ref{fig:HEWR} show the circumferential radius $R = r S = r \, \xi \, {\rm sech} (C \ln \xi) = e^t \, {\rm sech} (C \ln (e^t/r) )$ as a function of $r$ at two different times. The solution is homogeneous and so effectively incompressible. Close to the horizon (although in interpreting the word ``close,'' one should recall the units of $r$ and $t$), the radius increases linearly with $r^C$. At large $r$, $R$ declines with increasing $r$. For a fixed value of $t$, the maximum in $R$ occurs when $dR/dr = 0$. This maximum is most easily determined by setting $d(e^t/R)/d\xi = 0$, i.e., $d \cosh (C \ln \, \xi)/d \xi = 0$. This occurs when $\sinh (C \, \ln \xi) =0$, i.e., when $\ln \, \xi = 0$ or $\xi = 1$.

With $\xi = 1$, the maximum value of $R$, and the radius $r$ at which it occurs, both equal $e^t$.  Consequently the radius at which the circumferential radius is largest and the spatial geometry is a `two sphere', moves outward in time as $r_{\rm peak} = e^t$.

The lower panel shows the time dependence of the radius $R$ at the spatial coordinate $r=1$. The curvature in the plot reflects the different values of $H$. (The time scale for the evolution is of course very long.) 

For a fixed value of $r$, the maximum in $R$ occurs when $dR/dt = 0$. This is best evaluated by setting $d(r/R)/d\xi = 0$, i.e., $d \, [\cosh (C \ln \xi )/\xi] / d \xi = 0$ (note the appearance of the $\xi$ term in the denominator, compared to the expression above that maximizes $R(r)$ for a given $t$). This occurs when $\tanh (C \, \ln \xi ) = 1/C $, which is only slightly less than unity, so that the argument of the $\tanh$ function is large.

Using the high-argument approximation $\tanh x \simeq 1 - 2 e^{-2x}$ gives $e^{2 \, C \, \ln \, \xi} \simeq 2C/(C-1)$ or $\xi = (2C/(C-1))^{1/2C}$. Further using the identity ${\rm sech} \, x = \sqrt{1 - \tanh^2 x}$, the corresponding maximum value of $R$ is found to be $r \, (2C/(C-1))^{1/2C} \, \sqrt{C^2 - 1}/C $. For values of $C$ close to unity, the maximum value of $R \simeq 2$, and it occurs at $\xi \simeq \sqrt{2/(C-1)}$ or $t = \ln (\sqrt{2/(C-1)} \, r)$.

The time dependence may also be understood at fixed $r$ because  $\xi$ increases exponentially with $t$.  However $R= e^t \, {\rm sech} (C \, \ln \xi))$ eventually declines as $e^{(1-C)t}$ at very large $t$. Before this can happen however, the singularity reaches the chosen $r$ when $\xi=\xi_h=13.81$. Thus at $r=1$ this occurs at $t=2.62$ while the maximum doesn't pass until $t=2.65$ according to our estimate above.  Physically, the time dependence at every fixed $r$ reveals the relentless expansion of this cosmic singularity.

\subsection{Matter Pressure}

\begin{figure}[pht]
\includegraphics[width=0.99\linewidth]{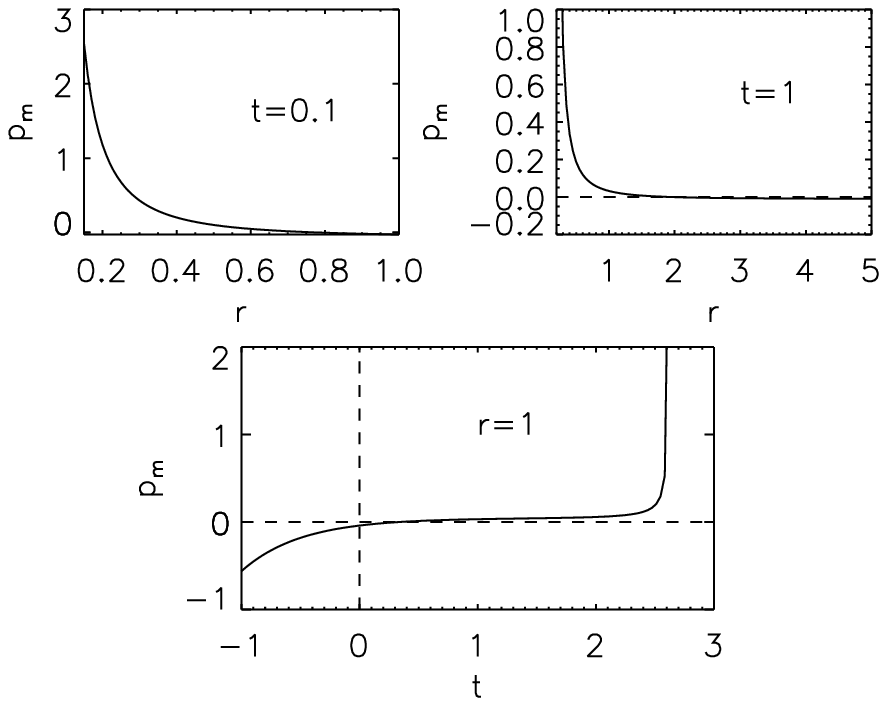}
\caption{At the upper left we show the matter pressure $p_m$ as a function of $r$ at $t = 0.1$.  The pressure goes to zero at about the location of our observer ($\xi_o = 0.78, r_o = 1.42$) and diverges positively at small $r$ near the horizon. At the upper right we plot the same quantity at the later time $t=1$. The divergence at the horizon ($r_h = 0.197$) is now much closer to our observer at $r_o=3.48$. In the lower panel we have plotted the pressure as a function of $t$ at $r=1$. In the future, at $t \approx 2.62$, the pressure diverges as the horizon overtakes this coordinate position. (Our observer is at $t = -0.25$.)}
\label{fig:HEWP}
\end{figure}

\begin{figure}[pht]
\includegraphics[width=0.95\linewidth]{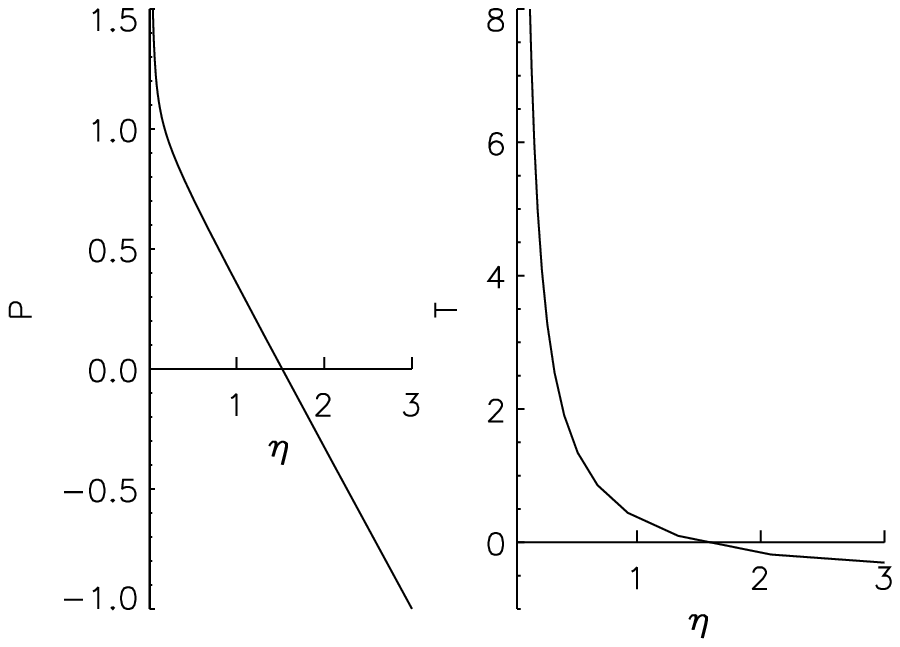}
\caption{Equation of state. {\it Left panel:} Variation of the pressure variable $P$ with density variable $\eta$. {\it Right panel:} Same information, but presented as a plot of ideal gas temperature $T$ vs $\eta$. \label{fig:PTrho}}\end{figure}

The most physically delicate element of the embedding cosmology is the pressure. Figure \ref{fig:HEWP} indicates that the pressure is small (positive or negative) almost everywhere, except at the horizon where it diverges positively. This divergence is evident at small $r$ at fixed $t$ and in the future at fixed $r$, as the apparent horizon passes. Given the homogeneous density, and approximating the co-moving plasma as an ideal gas, we see that this divergence indicates the presence of an infinite ideal gas temperature ($T \propto P/\eta$). Except for the singularity at the center of the system, there is no similar divergence in the absence of a horizon singularity, when $C<1$. 

In the left panel of Figure~\ref{fig:PTrho} we show the pressure variable $P$ as a function of the density variable $\eta$ (see Equation~(\ref{eq:mdensmpress})). This treats the equation of state as barotropic and shows an apparently un-physical pressure decline with increasing density. However this is misleading because, assuming an ideal gas equation of state (as in equation (\ref{eq:EOS}) below) the pressure, density, and temperature are each  functions of the self-similar variable $\xi$, and so all functions of one another. 

If we separate out the `temperature' (measured in units of $\overline{m} c^2/k$ by introducing the mean particle mass $\overline{m}$ and Boltzmann's constant $k$) and the density as in Equation~(\ref{eq:EOS}), we obtain the graph in the right panel of Figure~\ref{fig:PTrho}. This shows that it is the variation of effective temperature that is responsible for the behavior of the pressure. The temperature increases with decreasing density (reaching infinity at the singularity). It becomes negative at larger densities, which we can interpret as the onset of gravitational instability.  

Returning to the pressure in space and time with Figure~\ref{fig:HEWP}, we see that at a given time $t$, both $\xi$ and $P(\xi)$ decrease as $r$ increases. Since the density is homogeneous, the temperature cools with decreasing pressure, so that, just as in the standard $\Lambda-{\rm CDM}$ model, the cosmology outside the horizon will become transparent. 

To better understand the equation of state we can write it by eliminating the factor $1/(K\xi)^2$ in $P$ using $\eta$, both as written in equations (\ref{eq:cosmosol}). This gives the explicit equation of state in terms of the self-similar variable 

\be
p_m=\frac{\rho_m}{3}\left [ \frac{(3C-1)\xi^{2C}-(3C+1)}{(C+1)-(C-1) \, \xi^{2C}} \right ] \,\,\, ,\label{eq:EOS}
\ee
which we emphasize is imposed by the kinematic self-similar symmetry. The factor multiplying the density is the temperature as a function of the self-similar variable, assuming an ideal gas equation of state.

According to this equation of state, $p_m$ will decline smoothly (see Figures~\ref{fig:HEWP} and~\ref{fig:PTrho}) to the relativistic value $\rho_m/3$ when $\xi$ has declined to a value

\be
\xi_r=\left( \frac{2C+1} {2C-1} \right)^{1/2C} \,\,\, .
\label{eq:xiR}
\ee
Further, the pressure reaches zero at $\xi = \xi_T$, where
\be
 \xi_T = \left( \frac{3C+1} {3C-1} \right)^{1/2C} \,\,\, ; \label{eq:xit}
\ee
These points represent, for a fixed time $t$, the steady decline with distance $r$ of the gas temperature to zero. Each state is attained at smaller times $t$ for smaller $r$ values.

For the limiting case $C=1$, the explicit equation of state is

\be
p_m = \frac{\rho_m}{3} \, (\xi^2-2) = \frac{\rho_m}{3} \left (\frac{e^{2t}}{r^2} - 2 \right ) ,\label{eq:EOS1}
\ee
and $\xi_r=\sqrt{3}$, $\xi_T=\sqrt{2}$. 

The pressure continues its decline with $r$ to negative values at large $r$ (smaller $\xi$). Negative values of $P$ in the range $\xi<\xi_T$ suggest a correspondence with local gravitational collapse, i.e., with star formation. This phase begins at some value $\xi_S$, which is always greater than $1$ being infinitesimally close to $\xi_T$.  At each $r$, this is after the passage of the ``announcement'' signal at $\xi=1$. 

We have placed our observer at $\xi_o=0.78$ 
because $H$ is comparable to what is observed at this wave front. This suggests that the interesting region for cosmology is close to $1$ because $H$ soon becomes unreasonably large if $\xi_o$ is smaller.

It is worth recalling that the pressure is closely related to the geometric curvature in a homogeneous space-time. Assuming the stress tensor of an ideal ``fluid,'' as was done in \cite{HEW1983}, one has for the scalar curvature
\be
\Re=\frac{G}{c^2} \left ( \frac{3 \, p}{c^2} - \rho \right ) \,\,\, ,
\ee
where $p$ and $\rho$ are total quantities.

\subsection{Matter Mass}

\begin{figure}[pht]
\centering
\includegraphics[width=0.99\linewidth]{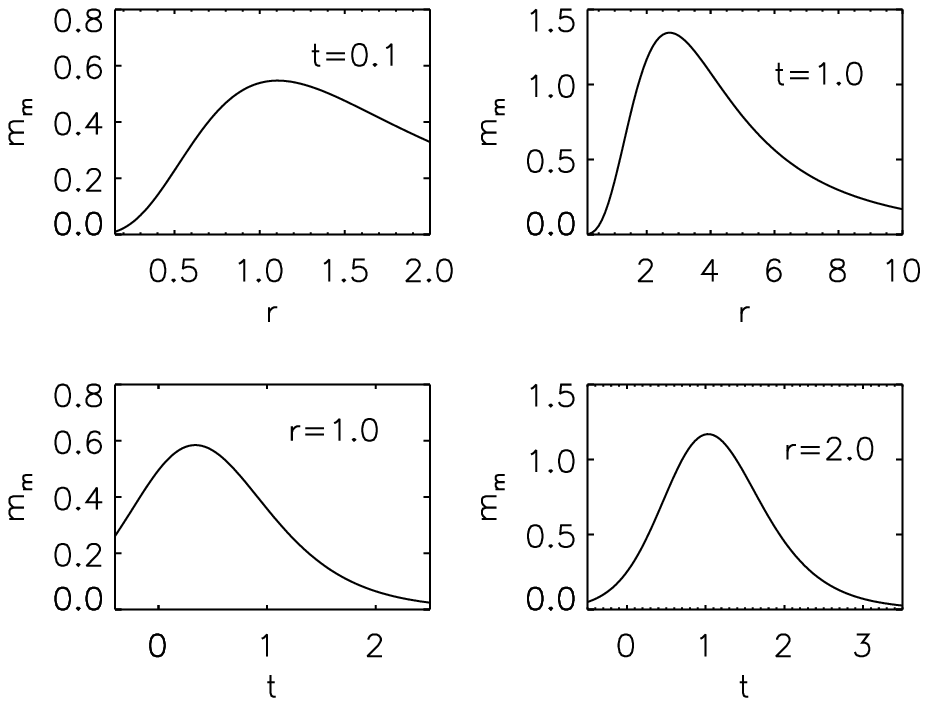}
\caption{The panel at the upper left shows the variation of the physical matter mass $m_m$ with $r$ at $t=0.1$. The singularity is at $r_h=0.08$ and our observer is at $r_o=1.417$. At upper right the behavior of $m_m$ is shown at the later time $t=1$, when $r_h=0.197$ and $r_o=3.48$. The maxima occur at $\xi=1$, corresponding to $r = e^t \simeq 1.1$ and $2.72$, respectively, with  maximum values of $e^t/2 \simeq 0.55$ and $1.38$, respectively (see text). At the lower left we show the dependence of $m_m$ on $t$ for $r=1$. The singularity arrives when $t \simeq 2.62$ and the observer is at $t=-0.25$. At the lower right we show the time dependence at $r=2$. The observer passes there at $t=0.445$ and the singularity arrives at $t=3.32$.}

\label{fig:HEWM}
\end{figure}

We can also look at the matter mass of the Universe. The matter mass and the vacuum mass satisfy the simple relations $m_m=(4\pi/3) \, \rho_m \, R^3$ and $m_v = (4\pi/3) \, \rho_v \, R^3$, respectively, because both densities are homogeneous (although $\rho_m$ is a function of $t$). We see that at a fixed $t$, both masses will vary with $r$ as $[R(r)]^3$. This yields a rise and fall of each mass, following the dependence of $R$ on $r$, as shown in Figure~\ref{fig:HEWR}. We illustrate the behavior in space and time in Figure~\ref{fig:HEWM}.

The variation in $r$ with $t$ at a fixed radius $r$ shows an initial rise of $m_m$, followed by a fall as the singularity compresses mass up until $\xi=1$; beyond this surface the mass declines. For a fixed value of $t$, the maximum in $m_m$ occurs when $dm_m/dr = 0$. This is most easily determined by setting $d(e^t/2 \, m_m)/d\xi = 0$, i.e., $d \cosh^3 (C \ln \, \xi)/d \xi = 0$. This occurs when $\sinh (C \, \ln \xi) =0$, i.e., $\ln \, \xi = 0$ or $\xi = 1$. With $\xi = 1$, the maximum value of $m_m$ is $e^t/2$ and the radius $r$ at which it occurs is $e^t$. 

Similarly, for a fixed value of $r$, the maximum in $m_m$ occurs when $dm_m/dt = 0$. This is best evaluated by setting $d(r/m_m)/d\xi = 0$, i.e., $d \, [\cosh^3 (C \ln \xi )/\xi] / d \xi = 0$, which occurs when $\tanh (C  \ln \xi ) = 1/3C$. Again using the high-argument approximation $\tanh x \simeq 1 - 2 e^{-2x}$ gives $e^{2 \, C \, \ln \, \xi} \simeq 6C/(3C-1)$ or $\xi = (6C/(3C-1))^{1/2C}$. Further using the identity ${\rm sech} \, x = \sqrt{1 - \tanh^2 x}$, the corresponding maximum value of $m_m$ is found to be $(r/2) \, (6C/(3C-1))^{1/2C} \, (9 \, C^2 - 1)^{3/2}/27 \, C^3$. For values of $C$ close to unity, the maximum value of $m_m \simeq ( 8 \, \sqrt{6}/27 ) \, r$, and it occurs at $\xi \simeq \sqrt{3}$ or $t \simeq \ln (\sqrt{3} \, r)$. These approximate values agree well with the exact numerical values in Figure~\ref{fig:HEWM}.  The maximum mass occurs very close to $P=\eta/3$.

The surface $\xi = 1$ is hence the limit of influence of the expanding singularity. In both upper panels we see the change in the $r$ mass distribution with time. In the lower panels we see the matter mass declining at $r$ as the apparent horizon passes.

\subsection{Singular Horizon}

\begin{figure}
\includegraphics[width=0.99\linewidth]{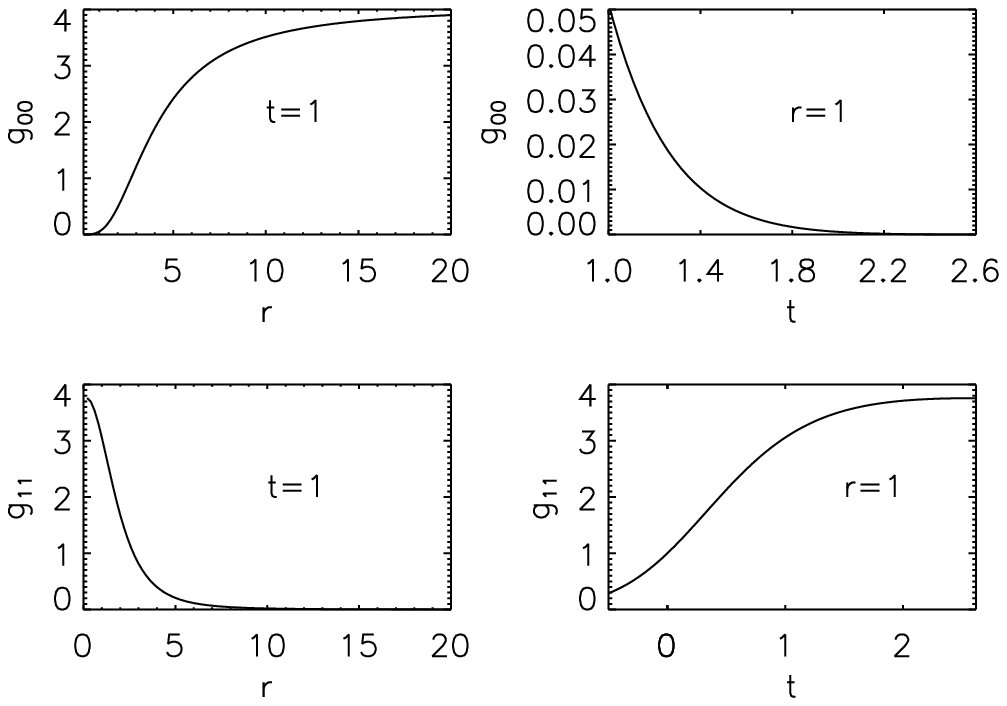}
\caption{At upper left we show the large scale evolution of $g_{00} \equiv e^\sigma$ as a function of $r$ at $t=1$. The coefficient rises from zero at the horizon $r_h = 0.197$ and approaches $(1+C)^2 \simeq 4$ at large $r$ relative to the horizon. (Our observer is at $r=3.78$.)  At upper right the time dependence of $g_{00}$ at $r=1$ is shown; the singularity arrives at $t=2.62$. (Our observer arrives at an earlier time, namely $t = -0.25$.) The lower left plot shows the metric coefficient $g_{11} \equiv e^\omega$ at $t=1$. The plot starts at the horizon value $r_h = 0.197$ and descends to a small, essentially constant value at large $r$. The plot at the lower right shows the time dependence of $g_{11}$ at $r=1$ from the passage of the observer to the passage of the apparent horizon.}
\label{fig:HEWG}
\end{figure}

In Figure~\ref{fig:HEWG} we look briefly at the approach to the  horizon, as measured by the behavior of $g_{00} \equiv e^\sigma$ as it approaches zero. The plot at the upper left shows that at $t=1$ the value of $g_{00} = e^\sigma$ rises from zero at the apparent horizon to approach a constant value at large $r$, which would define proper time far away from the horizon. The plot at the upper right shows that at $r=1$, the $t$ dependence declines steadily to zero as the singular horizon passes. 

The figure also includes the behavior of the metric coefficient $g_{11}=e^\omega$. The lower left plot, for $t=1$, shows that $g_{11}$ is well behaved at the horizon, although quite large. It diminishes at large distances to a small, slowly declining value. Space is therefore ``stretched'' relative to large $r$ near the horizon. At the lower right we show the time dependence of $g_{11}$ at $r=1$, from the passage of the observer at $t=-0.25$ to the passage of the apparent horizon at $t = 2.62$.

\subsection{Lorentz Factor and Four Velocity}
\label{sec:lorentz-factor}

 The plots of $\Gamma^2$ show clearly the significance of $\xi=1$, with $\Gamma$ vanishing there because of our imposed separation of the energy equation, so that $\Gamma^2=1-M/S\equiv 1-2 \, Gm_m/R c^2$. Therefore at $\xi=1$ (where $\Gamma^2=0$) $R$ has the Schwarzschild horizon value  namely $ 2 \, Gm_m/c^2$. The radial light velocity is equal to $\pm 1/(CK)$ at $\xi=1$ and becomes equal to unity with $K=1/C$. This sets the light speed as the signal announcing the approach of the singularity.

\begin{figure}
\centering
\includegraphics[width=0.99\linewidth]{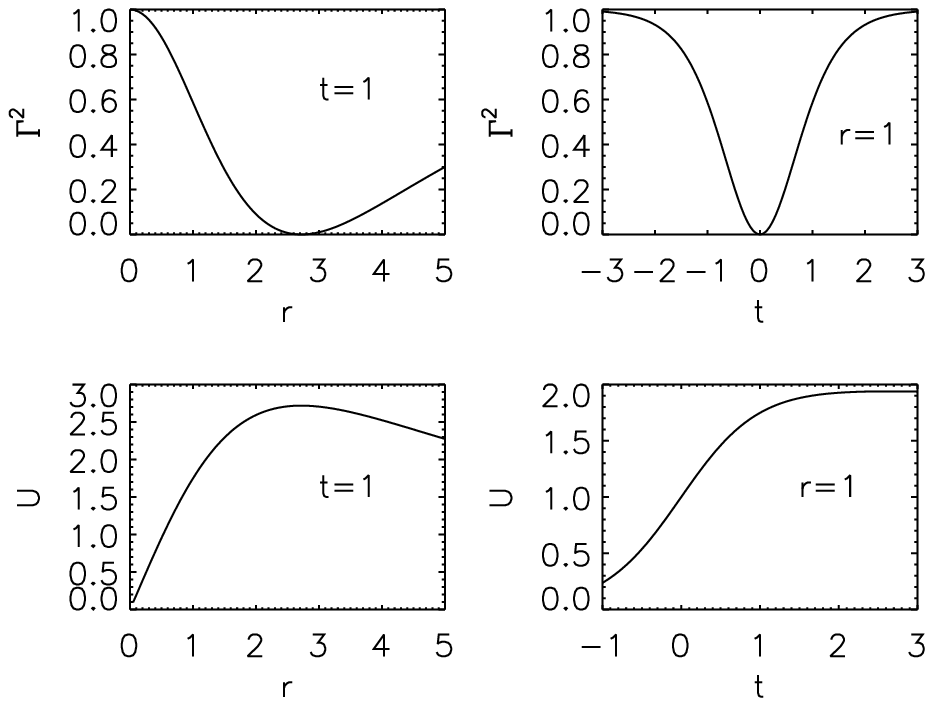}
\caption{In the upper left panel we plot the square of the Lorentz factor $\Gamma^2$ as a function of $r$ for $t=1$. $\Gamma$ reaches zero at $r=e$ which corresponds to $\xi=1$. (Our observer is at $r_o = 3.48$.) On the right we show the time dependence at $r=1$; $\Gamma$ now reaches zero at $t=0$, which again corresponds to $\xi=1$. (Our observer now is at $r_o=-0.25$.) At lower left we show the spatial dependence of $U$ for $t=1$; the maximum in $U$ occurs at $\xi = 1$, i.e., at $r = e$. (The observer is again at $r_o = 3.48$.) At the lower right the time dependence of $U$ on $t$ is shown for $r = 1$. The maximum occurs at $\xi=\xi_h=13.81$ which happens at $t = 2.62$. (Our observer is at $t_o = -0.25$ and the surface $\xi=1$ is at $t = 0$.) }
\label{fig:HEWfigUgamma}
\end{figure}

It is also seen in Figure~\ref{fig:HEWfigUgamma} that $\xi=1$ locates the maximum four radial velocity in space at each fixed time (see lower left in the figure). This reflects that $\partial_rU=0$ at $\xi = 1$. We note also that $\partial_t U =0$ at $\xi_h$, so that at fixed $r$, the maximum $U$ occurs at $t_h = \ln (r \, \xi_h)$, as at the lower right in the figure. We note that the maximum values of $U$ can be greater than unity.

This behavior is similar to the maximum in $S$ at the horizon $\xi_h$, which defines the maximum in $R$ at fixed $r$ (Figure~\ref{fig:HEWR}). In fact we can infer the behavior of $U$ from Equation~(\ref{eq:fourvel}) because $U = R$. 

The geometrical nature of $\xi_h$ remains a little mysterious. We present a clarification in the appendix and section~\ref{sect:LC}.  The physical  significance of $\xi=1$ is that of  a precursor wave representing the limit of influence of the expanding singularity in the expanding cosmology. This surface is a spatial ``two sphere'' of maximum four velocity at each fixed time $t$. The radial light speed is equal to unity ($K=1/C$) and the circumferential radius equal to the Schwarzschild radius.

\bigskip

We conclude this section by looking at quantities evaluated at the  horizon as approached from the `outside' ($r>e^t/\xi_h$). 

Using the expressions in Equation~(\ref{eq:cosmosol}), we find that $\tanh (C \, \ln \xi_h) = 1/C$, so that $R(\xi_h) = Ke^t \, \sqrt{1-1/C^2}$, where we recall that the unit of length is $\sqrt{3/\Lambda}$ and that of time is $\sqrt{3/\Lambda \, c^2}$.

The matter density, in units of $\rho_v$, is $(3/8\pi) \, e^{-2t}$ and the matter pressure is infinite. The matter mass in units of $\mu$ (Equation~(\ref{eq:value_of_mu})) is $m_m = ((1-1/C^2)^{3/2}/2) \, e^t$ and the vacuum mass $m_v = 4 \pi R^3 \, \rho_v/3$, in the same units, is $((1-1/C^2)^{3/2}/2) \, e^{3t}$. The metric coefficients are $g_{00} = 0$ and $g_{11} = \xi_h^2 \, (C^2-1)$.

Because of the magnitude of the mass unit $\mu$, with $C=1.01$ we need $e^{t_h} = 6.4 \times 10^{-15}$ to restrict the black hole mass to $10^6 M_\odot$ while holding $\xi=\xi_h =13.8$. This implies that $r_h=e^{t_h}/13.8\simeq 4.6 \times 10^{-16}$ in terms of the length unit. This is $\simeq 7.6 \times 10^{12}$~cm. 
For comparison, the Schwarzchild radius for a $10^6 M_\odot$ black hole is 25 times smaller, $3 \times 10^{11}$~cm. 
If, however, $C$ takes on the limiting value of unity, then $\xi_h \rightarrow \infty$, and the matter mass within the singular region $m_m = (r_h/2) \, \xi_h \, {\rm sech}^3 (\ln \xi_h)$ (equations~(\ref{eq:cosmosol})) goes to zero as $4 \, r_h/\xi_h^2$.

\section{Limiting Case}\label{sect:LC}

When $C=1$, the singular horizon and a central singularity are together at $\xi_h=\infty$ (see Equation~(\ref{eq:xih})). This value will always be achieved at $r=0$ for all finite $t$, which might suggest that the singularity is isolated there. 

However the central feature of any of these solutions is the surface $\xi = {\rm constant}$. This quantity becomes infinite at any finite $r$ as $t \rightarrow \infty$, which therefore implies (in contradiction to the isolation at $r=0$) that at infinite time all $r$ encounter the singularity at $\xi \rightarrow \infty$. These statements are confirmed in the appendix, where we find the singularity at $i^+$ and at $r=0$ to be combined. Lines of constant $r$ (i.e. $\rho$ in the appendix) do intersect the singularity at $i^+$. This is an example of a space-like singularity, but at $i^+$.

Although we have discussed physical behavior for the possible interest in cosmology, we emphasize that the truly rigorous conformal picture in this paper is for the case $C=1$. The appendix should therefore be regarded as an essential part of the text, which reveals the technical detail.

Guided by the non-singular case $C<1$ and the rigorous case $C=1$ (see appendix), we suspect that the situation for $C>1$ is as shown in Figure~\ref{fig:conformal}. These are only schematic diagrams that are not conformal, but they are well founded For the limiting case $C=1$ and the cosmological case $C<1$ \citep[e.g.,][p. 726]{Pen2004}. We give a brief justification of these diagrams below.

Although global solutions for the null geodesics are not available when $C\ne1$, we can use piece-wise solutions for $t<0$ and $t>0$ separately. These approximations work best for $C$ large and $r$ not too small. They yield our justification for the schematics in figure (\ref{fig:conformal}).

For $t>0$ the first term in equation (\ref{eq:null_path_0}) is dominant. This integrates to give
\bea
r^C&=&C\,u-\frac{C}{2}\,e^{(C-1)t},\label{eq:outt+}\\
r^C&=& C\,v+\frac{C}{2}\,e^{(C-1)t},\label{eq:int+}
\eea
for ingoing and outgoing geodesics respectively. (We must interchange the $+$ and $-$, because the $\mp$ assumes that the following factor is negative.) The quantities $v$ and $u$ are arbitrary null coordinate parameters. 

The ingoing ray stops at finite $t=(\ln{2\,u})/(C-1)$ where $r=0$ and $\xi=\infty$. The outgoing ray continues to $i^+$ where it encounters the singularity according to $r_h^C=e^{Ct}/\xi_h^C$. Should $v<0$, the ray can begin at $r=0$ at time $\ln{2|v|}/(C-1)$.

When $t<0$, the second term in equation (\ref{eq:null_path_0}) dominates, which is properly negative. This integrates to
\bea
r^C&=& \frac{2}{C\,v+C\,e^{-(C+1)t}},\label{eq:outt-}\\
r^C&=& \frac{2}{C\,u-C\,e^{-(C+1)t}},\label{eq:int-}
\eea
for outgoing and ingoing rays respectively. All of the outgoing rays begin at $r=0$ at $i^-$. 

Should $v<0$, it appears that the outgoing ray can reach infinity at $|t|=\ln{|v|}/(C+1)$. However the singularity is at $r_h^C=e^{-C|t|}/\xi_h^C$ and the outgoing radius coincides with this value at $|t|=\ln{(2\xi_h^C/C\,e^{C|t|}+|v|)}/(C+1)$. Being a larger negative time then the time to reach infinity, this happens before the null ray reaches $i_o$.

The ingoing rays ($u>0$) start at infinity at time $|t|=\ln{u}/(C+1)$ and continues inward to some finite value.

These comments agree with the schematic diagrams in figure (\ref{fig:conformal}) that are not conformally transformed. The special case $C=1$ is transformed conformally in Figure~\ref{fig:strconform}, which agrees topologically with the schematic version. When $C<1$ the situation is fundamentally different, because there is no propagating singular horizon.

\bigskip

\section{Discussion}

In view of the recent interest in black holes with non-flat asymptotic boundaries, we have studied a known solution of the general relativity field equations in a new parameter space. The space-time is space-time scale-invariant, spherically symmetric, non stationary, matter homogeneous, and possesses  a positive cosmological constant $\Lambda$. This constant and the speed of light $c$ together define space and time scales that are cosmological in size. Changes in dimensionless space and time variables of order unity can therefore span cosmological scales.

We have extended the solution of \cite{HEW1983} so that the key parameter $C$ is now permitted to be $\ge 1$, and we have expressed this parameter in terms of physical quantities (Equations~(\ref{eq:physicalC}) and~(\ref{eq:beta-definition})). The expanding horizon is unusual because it is singular with an infinite pressure and a correspondingly infinite scalar curvature. When $C=1$ the horizon is at $\xi = \xi_h \rightarrow \infty$. This places it  at $r = 0$ and $t = i^+$ (so it engulfs all $r$ as $t \rightarrow \infty$). This case is rigorously interpreted conformally in the appendix of this paper. The solution for $C>1$ is exact, but the interpretation does not result in a precise Penrose-Carter diagram.

We have often referred to our object as an expanding `black hole' for brevity, but this pressure and curvature singularity at the expanding horizon clearly renders it a fundamentally new type of object when $C>1$. It is a growing, `hot,' curvature singularity that nevertheless is not `naked.' 

Although referring to the schematic conformal diagrams may be more helpful, a simple plot of $\xi_h\ne 0$ in a $(t$-$r)$ plane shows that light emitted tangentially from the singularity at some $(t$,$r)$ arrives at infinity after the singularity does. (This is confirmed by a direct calculation of the null path.) Hence the singularity is never visible to an observer outside the horizon when $C>1$.

Starting from a central point, an expanding, singular, apparent horizon eventually engulfs the whole of the space-time as a kind of wave front in an incompressible medium. However there is a preceding  null surface ($\xi=1$) that separates the part of the space-time manifold that is already affected by the expanding horizon, from that not yet affected.

The infinite pressure at the apparent horizon, coupled with the finite density there, also implies (for matter with an ideal equation of state) that the apparent horizon is a sphere of very high temperature, i.e., a ``sphere of fire,'' as illustrated in Figure~\ref{fig:HEWP}. This divergence of pressure and temperature also coincides with a curvature singularity. 

We note also, based on Figure~\ref{fig:HEWP}, that the matter pressure can be \emph{negative} at early times before the passage of the `warning' light front at the observer's location. This would assist the negative vacuum pressure in accelerating the cosmology at early times (and slow it down at later times). Negative pressure might be associated with local gravitational collapse, well before the onset of virial equilibrium in galaxies. 

The visible aspects of the hot early Universe would be due to hot plasma inside the point where $\xi=\xi_T$ that is becoming transparent. This is much as is the case for the standard current cosmology. 

An observer situated in the unaffected space-time, who calculates the Hubble-Lema\^itre constant using coordinate time, can observe currently favored values.  A calculation of $H$ using proper time gives a constant value $\sqrt{\Lambda c^2/3} = 56$~km~s$^{-1}$~Mpc$^{-1}$.

The radial light velocity is zero at the horizon, but the horizon hyper surface normal is not null unless $C=1$. In the latter case  we have seen that the behavior is an expanding black hole in a $\Lambda$-CDM cosmology. This was our original objective.

The generalized Lorentz factor $\Gamma$ and the comoving  four velocity $U$ are studied in Figure~\ref{fig:HEWfigUgamma}. The Lorentz factor vanishes at $\xi = 1$, which identifies this light front as a Schwarzschild surface satisfying $R = 2 \, G \, m_m/c^2$. The four velocity is permitted to be positive or negative.

In conclusion we have described a space-time with an infinitely `hot' horizon that is expanding into a surrounding cosmology. This is a much more dynamic object than a static black hole sitting in an asymptotically flat space time. It resembles one interpretation of the McVittie metric (\citep{LA2011}). The case $C=1$ describes a singular black hole horizon that becomes a space-like future event horizon.  The solution inevitably develops the `firewall' behavior at the horizon.   

\backmatter

\bmhead{Acknowledgements}
RNH thanks his wife Professor Judith Irwin for her quiet encouragement.

\begin{appendices}

\section{Conformal Diagrams}\label{sec:conformal_diagrams}

\begin{figure}
\centering
\includegraphics[width=0.9\linewidth]{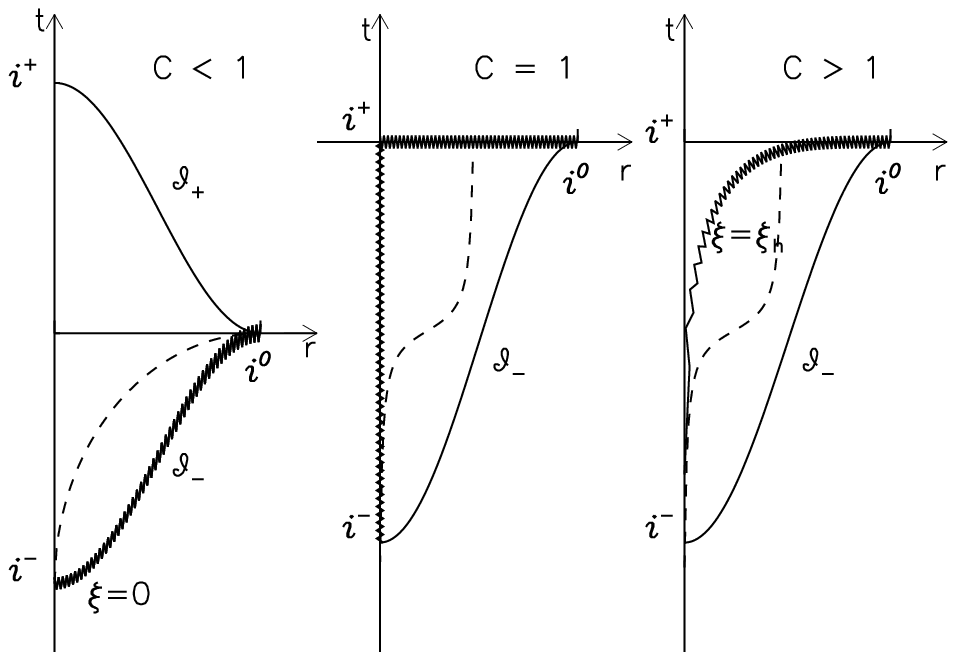}
\caption{Schematic conformal diagrams for the three pertinent values of the parameter $C$; in each case the horizon singularity is depicted by a zig-zag line and an illustrative null ray is shown as a dashed line. We use $(t,r)$ coordinates to locate the singularity accurately. Hence the light paths are distorted. The exterior region is below and to the right of the singularity locus in the second and third sketches. For $C \ge 1$ that ray eventually (and necessarily) encounters the singularity at finite values of $r$ and $t$.}
\label{fig:conformal}
\end{figure}

Figure~\ref{fig:conformal} shows `schematic' conformal diagrams for the cases $C<1$ (the original cosmology of \cite{HEW1983}), $C>1$ (where a horizon exists at finite values of $t$ and $r$), and $C=1$ (the limiting case discussed in Section~\ref{sect:LC}). In the limiting case $C=1$, we can be more quantitative. For this value of $C$, Equations~(\ref{eq:cosmosol}) give
\bea
e^\sigma  = & \,\,  (\tanh (\ln \xi) ) -1 )^2 = \frac{4}{(\xi^2+1)^2}  \, \,\, ;\cr
e^\omega  =  & \!\!\!\!\!\! \xi^2 \, {\rm sech}^2 (\ln \xi)  = \frac{4 \, \xi^4}{(\xi^2+1)^2} \,\,\, ,
\eea
so that the part of the metric corresponding to radial motion is
\be\label{eq:radial-metric-c=1}
ds^2 = \frac{4}{(\xi^2+1)^2} \,  \left ( dt^2 - e^{4t} \, \frac{dr^2}{r^4} \right ) \,\,\, .
\ee
With the change of variable $\rho = 1/r$, this becomes
\be\label{eq:radial-metric-rho}
ds^2 = \frac{4 \, e^{4t}}{(\xi^2+1)^2} \,  (e^{-4t} \, dt^2 - d\rho^2) \,\,\, .
\ee
Radial null geodesics satisfy
\be\label{eq:radial_null_geodesic}
\pm \, e^{-2t} \, dt = d\rho \,\,\, ,
\ee
which suggests that we define a new time variable $\tau$ through
\be
d\tau=e^{-2t}\,dt \,\,\, .\label{eq:nullpathtau}
\ee
This integrates to
\be
\tau=\frac{1}{2} \, \left ( 1-e^{-2t} \right ) \, ; \qquad t = \ln \, \left  ( \frac{1}{\sqrt{1-2\tau}} \right ) \,\,\, , \label{eq:timetrans}
\ee
where the constant of integration has been chosen so that $\tau$ and $t$ vanish together at $t=0$, and they are also at negative infinity together. $t=+\infty$ corresponds to $\tau=1/2$.

Using Equation~(\ref{eq:timetrans}), the self-similar variable
\be\label{eq:xi-rho-tau}
\xi \equiv \frac{e^t}{r} = \frac{\rho}{\sqrt{1-2\tau}} \,\,\, ,
\ee
and so the metric~(\ref{eq:radial-metric-rho}) becomes
\be
ds^2 = \frac{4}{(1 - 2 \tau + \rho^2 )^2} \, \,  (d \tau^2 - d\rho^2) \,\,\, .
\ee
We can now make the usual change to ingoing and outgoing Eddington-Finkelstein coordinates $u = \tau - \rho, v = \tau + \rho$, respectively, and then, via the compactions ${\cal U} = \tan^{-1} u, {\cal V} = \tan^{-1} v$, to Kruskal-Szekeres coordinates ${\cal T} = {\cal V} + {\cal U}; {\cal R} = {\cal V} - {\cal U}$. This gives, after some reduction,
\be
ds^2 \! = \! \frac{4 \, (\cos {\cal T} \! + \! \cos {\cal R})^2 \, (d{\cal T}^2 - d{\cal R}^2)}{ \left [ (\cos {\cal T} \!\! + \!\! \cos {\cal R})^2 \! + \! \sin^2 {\cal R} \! - \! 2 \sin {\cal T} (\cos {\cal T} \!\! + \!\! \cos {\cal R}) \right ]^2} \,\,\, .
\ee

\begin{figure}
\centering
\includegraphics[width=0.8\textwidth]{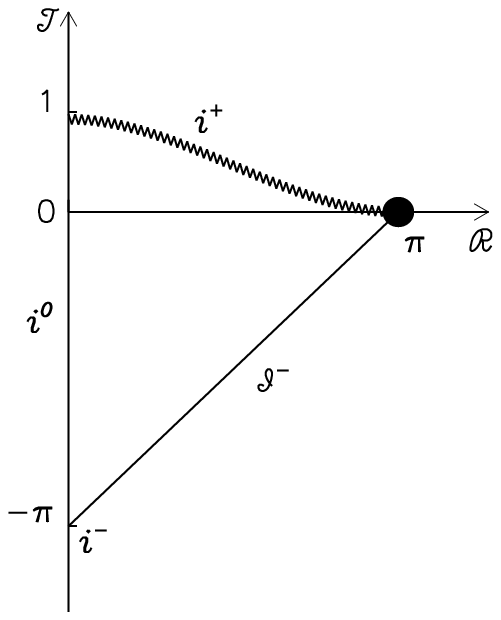}
\caption{Strict conformal diagram of the $C=1$ solution in $({\cal R}, {\cal T})$ space, with the horizon singularity shown as a zig-zag line.}
\label{fig:strconform}
\end{figure}

Figure~\ref{fig:strconform} shows the conformal diagram in $({\cal R}, {\cal T})$ space.  Radial null geodesics are defined by $d{\cal T} = \pm \, d{\cal R}$, i.e., the usual $45^\circ$ lines. Outgoing rays (increasing $r$, increasing $t$) correspond to decreasing ${\cal R}$ and increasing ${\cal T}$ (i.e., they point toward the top left), while ingoing rays point to the top right. Lines of constant $r$ and constant $t$ correspond to lines of constant $\rho$ and constant $\tau$, respectively, which can be written
\be
\frac{\sin {\cal R}}{\cos {\cal T} \!\! + \!\! \cos {\cal R}} = {\rm const} \, ; \quad \frac{\sin {\cal T}}{\cos {\cal T} \!\! + \!\! \cos {\cal R}} = {\rm const} \,\,\, ,
\ee
respectively. Points with constant $\xi$ are defined by the loci
\be\label{x-R-T}
\xi = \frac{\sin {\cal R}}{\sqrt{(\cos {\cal T}  \!\! + \!\! \cos {\cal R}) (\cos {\cal T} \!\! + \!\! \cos {\cal R} - 2 \sin {\cal T})}} \,\,\, ,
\ee
so that $\xi \rightarrow \infty$ when $2 \, \sin {\cal T} - \cos {\cal T} = \cos {\cal R}$; future timelike infinity $i^+$ also corresponds to this locus. 

At ${\cal R} = \pi$, $i^+$ is at ${\cal T} = 0$; at ${\cal R} = \pi/4$, $i^+$ is at ${\cal T} = \pi/4$, and at ${\cal R} = 0$ it is at ${\cal T} = 2 \ \tan^{-1} (1/2) \simeq 0.93$. 
Past time like infinity $i^-$ corresponds (Equation~(\ref{eq:timetrans})) to $\tau = -\infty$ and hence to $u=v = -\infty$, ${\cal U} = {\cal V} = - \pi/2$, and so to the point $({\cal R} = 0, {\cal T} = -\pi)$. 

The inward null line $\mathcal{I}^-$ corresponds to ${\cal R} - {\cal T} = \pi$, while the outward null line $\mathcal{I}^+$ corresponds to ${\cal R} + {\cal T} =\pi$, which lies outside the domain corresponding to points in the observable universe and so is not shown. 

Finally, as $r \rightarrow \infty$ ($\rho=0$) at finite $t$, $u = v = \tau$, ${\cal U} = {\cal V} = \tan^{-1} \, \tau$ and so $({\cal T} = 2 \, \tan^{-1} \, ((1 - e^{-2t})/2 ), {\cal R} = 0)$. Thus space like infinity $i^0$ corresponds to a vertical line segment on the ${\cal T}$ axis, extending from ${\cal T} = -\pi$ to the singularity at ${\cal T} = 2 \, \tan^{-1} (1/2)$.

The point $({\cal R} = \pi, {\cal T} = 0)$ corresponds to ${\cal V} = \pi/2, \,{\cal U}=-\pi/2$, or $v = \infty, u = -\infty$, i.e., $\rho=\infty$ or $r=0$. Thus the central singularity shown schematically (\ref{fig:conformal}) is reduced to this point. All in going light-like rays eventually intersect the (combined) singularity $\xi_h = \infty \equiv i^+$, while outgoing rays intersect either the singularity or space like infinity ($i^0$) on the $T$ axis.

Although we have demonstrated a quantitatively accurate conformal diagram in these coordinates only for the limiting case $C=1$, the conformal diagrams for $C = 1^+$ (cf. Figure~\ref{fig:conformal}) will be similar.  Obtaining `strict' conformal diagrams for these cases is outside the scope of the present work.

\end{appendices}


\begin{thebibliography}{20}
\bibitem[\protect\citeauthoryear{Farrah et al.}{2023a}]{FCZT2023} Farrah, D., Croker, K.S., Zevin, M., Tarl\'e, G., Faraoni, V., Petty, S., and thirteen more 2023, Ap. J., {944L}, {31}

\bibitem[\protect\citeauthoryear{Croker et al.}{2023}]{CFPT2023} Croker, K.S., Farrah, D., Petty, S., Tarl\'e, G., Zevin, M., Hatziminaoglou, E., and eleven more 2023, Bulletin of the American Astronomical Society, Vol. 55, 24141904C


\bibitem[\protect\citeauthoryear{McVittie}{1933}]{McV1933} McVittie, G.C. 1933, MNRAS, {93}, 325

\bibitem[\protect\citeauthoryear{McVittie}{1967}]{McV1967} McVittie, G.C. 1967,{\it Gravitational motions of collapse or of expansion in general relativity}, Ann.Inst,H.Poincar\'e, A6,1

\bibitem[\protect\citeauthoryear{Misner}{1969}]{M1969} Misner, C.W. 1969, Brandeis U. Summer Institute in Theoretical Physics 1968, Astrophysics and General Relativity v1 (Gordon and Breach: New York), pp 113-188

\bibitem[\protect\citeauthoryear{Lake \& Abdelqader}{2011}]{LA2011} Lake K., \& Abdelqader, M. 2011, Phys. Rev. D, {84}, {044045}

\bibitem[\protect\citeauthoryear{Henriksen, Emslie \& Wesson}{1983}]{HEW1983} Henriksen, R.N., Emslie, A.G. \& Wesson, P.S. 1983, Phys. Rev. D, {27}, {6}, {1219}

\bibitem[\protect\citeauthoryear{Carter \& Henriksen}{1991}]{CH1991} Carter, B., \& Henriksen, R. N. 1991, J. Math. Phys., {32 (10)}, {2580}


\bibitem[\protect\citeauthoryear{Henriksen}{2015}]{H2015} Henriksen, R. N. 2015, {\it Scale Invariance: Self-Similarity of the Physical World}, Wiley-VCH, 69469 Weinheim, Germany

\bibitem[\protect\citeauthoryear{Dias et al.}{2023}]{DGSB2023} Dias, O.J.C., Gibbons, G.W., Santos, J.E., \& Benson, W. 2023, Phys. Rev. Lett., {131}, {131401}

\bibitem[\protect\citeauthoryear{Islam}{1985}]{Islam1985} Islam, J. N. 1985, {\it Rotating fields in general relativity} (Cambridge University Press: Cambridge)

\bibitem[\protect\citeauthoryear{Landau and Lifshitz}{1975}]{LL1975} Landau, L. D., \& Lifshitz, E.M. 1975, {\it Classical Theory of Fields} (Pergamon Press: Oxford)

\bibitem[\protect\citeauthoryear{Penrose}{2004}]{Pen2004} Penrose, Roger, 2004,{\it The Road to Reality: A Complete Guide to the Laws of the Universe} (Alfred A. Knopf, Random House, New York)

\bibitem[\protect\citeauthoryear{Kramer et al.}{1980}]{KSMH1980}Kramer, D., Stephani,H., MacCallum,M. \& Herlt, E. 1980,{\it Exact Solutions of Einstein's field equations} (Cambridge Monographs on Mathematical Physics, Cambridge,U.K.)


\end{thebibliography}
\end{document}